\documentstyle[12pt]{article}
\RequirePackage{graphicx}
\begin{document}

\sloppy

\title{On generalized probabilities:\\
correlation polytopes for automaton logic and generalized urn models,
extensions of quantum mechanics and parameter cheats}
\author{Karl Svozil\\
 {\small Institut f\"ur Theoretische Physik, University of Technology Vienna }     \\
  {\small Wiedner Hauptstra\ss e 8-10/136,}
  {\small A-1040 Vienna, Austria   }            \\
  {\small e-mail: svozil@tuwien.ac.at}}
\date{ }
\maketitle


\begin{abstract}
Three extensions and reinterpretations of nonclassical probabilities are reviewed.
(i) We propose to generalize the probability axiom of quantum mechanics
to self-adjoint positive operators of  trace one.
Furthermore, we discuss the Cartesian and polar decomposition
of arbitrary normal operators and the possibility to operationalize the corresponding observables.
Thereby we review and emphasize the use of observables which maximally represent the context.
(ii) In the second part, we discuss Pitowsky polytopes for automaton logic
as well as for generalized urn models
and evaluate methods to find the resulting Boole-Bell type (in)equalities.
(iii) Finally, so-called ``parameter cheats'' are introduced, whereby parameters
are transformed bijectively and nonlinearly in such a way that classical systems
mimic quantum correlations and {\it vice versa}.
It is even possible to introduce parameter cheats which violate the Boole-Bell type
inequalities stronger than quantum ones, thereby trespassing
the Tsirelson limit. The price to be paid is nonuniformity.
\end{abstract}

\section{Hilbert space extensions}
Since Planck's introduction of the quantum  100 years ago \cite{planck:1901-v},
quantum mechanics has developed into a fantastically successful theory
which appears to be stronger than ever.
Despite its obvious relevance and gratifying predictive power, the question of how
to proceed
to theories beyond the quantum still remains not totally unjustified and is asked
by eminent and prominent researchers in the area \cite{greenberger-talk-99,shimony-talk-2000}.

Indeed, from an informal, historic perspective, it is quite likely that there will be
a successor theory of quantum mechanics of some sorts
which will extend this theory in many, hitherto unknown, aspects.
Such a theory might even be more mindboggling than quantum theory; or it may be based on
simple concepts such as information \cite{zeil-99};
or it may be a reinterpretation of the standard quantum formalism
in another set theoretic context \cite{pitowsky-82,clifton:99}.

A very radical and original approach to nonclassicality has been investigated by John Harding
 who argues that ``absolutely none of the structure of a Hilbert space
is necessary to produce an orthomodular poset. [[...Rather]]
it is a consequence of arithmetical properties of relations''  \cite{harding-96}.
In what follows, a much more humble extension is pursued which respects the established
framework of quantum mechanics.

Von Neumann's Hilbert space formalism \cite{v-neumann-49} of quantum theory
is extended
by considering more general forms of operators as proper realizations of physical observables.
From the point of view of vector space theory, these extensions reflect well-known properties of the
algebraic structures arising in quantum mechanics \cite{halmos-vs,svozil-ql}.
It may nevertheless be worthwhile to review them for a proper understanding of
the underlying physics.

Let us from now on consider finite dimensional Hilbert spaces.
The quantum  probability $P(\psi , A)$ of a proposition $A$ given a state $\psi$
is usually introduced as
the trace of the product of the state operators $\rho_\psi $
and the projection operator $E_A$; i.e.,
$P(\psi ,A)={\rm Tr}(\rho_\psi  E_A)$.
(This ``axiom'' of quantum probability has been derived
for Hilbert spaces of dimension larger than two
from reasonable basic assumptions by Gleason
\cite{Gleason,r:dvur-93}.)
The state operator $\rho_\psi$ must be
(i)  self-adjoint; i.e., $\rho_\psi =\rho_\psi ^\dagger$,
(ii) positive; i.e., $(\rho_\psi  x,x) =\langle x\mid \rho_\psi  \mid x\rangle \ge 0$ for all $x$, and
(iii) of trace one; i.e., ${\rm Tr} (\rho_\psi )=1$.
Since one criterion for a pure state is its idempotence; i.e., $\rho_\varphi \rho_\varphi =\rho_\varphi$,
one way to interpret $E_A$ is a measurement apparatus in a pure state $\rho_\varphi= E_A$.
But while pure states can be interpreted as a system being in a given property,
not every state is pure and thus corresponds to a projection.

We propose here to generalize quantum probabilities to properties corresponding
also to nonpure states, such that the general form of quantum probabilities can be
written as
\begin{equation}
P(\psi ,\varphi)={\rm Tr}(\rho_\psi  \rho_\varphi),
\end{equation}
where again we require that $\rho_\varphi$ is self-adjoint, positive and of trace one.
One immediate advantage is the equal treatment of the object and measurement apparatus.
They appear interchangeably, stressing the conventionality of the measurement process
\cite{svozil-2000interface}.

One property of the extended probability measure is its positive definiteness and boundedness;
i.e., $0\le P(\psi ,\varphi) \le 1$. The former
bound follows from positivity.
The latter bound by $1$ can be easily
proved for finite dimensions by making a unitary basis transformation such that $\rho_\varphi$ (or $\rho_\psi$) is
diagonal.

A very simple example of an extended probability is the case of the total ignorance of the
state of the measurement apparatus as well as of the measured system.
Take $n$ nondegenerate possible outcomes for
the apparatus and the state, then $\rho_\psi=\rho_\varphi={\bf 1}/n=(1/ n)\,{\rm diag}(1,1,\ldots, 1)$,
and the probability to find any combination thereof is
$P(\psi ,\varphi)=1/n^2$.
The extended probability reduces to the standard form
if one assumes total knowledge of the state
of the measurement apparatus, since then $\rho_\varphi$ is pure and thus a projection.


Another well known fact is the Cartesian and polar decomposition of an arbitrary
operator $A$ into operators $B,C$ and $D,E$ such that
\begin{eqnarray}
A&=&B+iC = DE,  \nonumber \\
B&=&{A+A^\dagger \over 2}, \quad C={A-A^\dagger \over 2i},\nonumber  \\
E&=&\sqrt{A^\dagger A},\quad D={AE^{-1}},\nonumber
\end{eqnarray}
where $B,C$ are self-adjoint, $E$ is positive and $D$ is unitary (i.e., an isometry).
The last two equations are for invertible operators $A$.
These are just the matrix equivalents of the decompositions of complex numbers.

If $A$ is a {\em normal} operator;
i.e.,  $AA^\dagger=A^\dagger A$, then
$B$ and $C$ commute (i.e., $[B,C]=0$)
and are thus co-measurable.
(All unitary and self-adjoint operators are normal.)
In this case, also the operators of the polar decomposition $D$ and $E$
are unique and commute;
i.e.,    $[D,E]=0$, and are thus co-measurable.
We have thus reduced the issue of operationalizability of normal operators
to the self-adjoint case; an issue which has been solved positively
\cite{rzbb}.

Hence,  normal operators are  operationalizable
either by a simultaneous measurement of the summands in the Cartesian decomposition or
or of the factors in a polar decomposition
(cf. also \cite{neumaier-pr,appleby-pr}).
Indeed, all operators are ``measurable''
if one assumes EPR's elements of counterfactual
physical reality   \cite[p. 108f]{svozil-ql}.
In this case, one makes use of the Cartesian decomposition, where $B$ and $C$
not necessarily
can be diagonalized simultaneously and thus need not commute.
Nevertheless, one may devise a singlet state of two particles with respect to the
observables $B$ and $C$, and measure $B$ on one particle and $C$ on the other one.

As an example for the case of a normal operator which is neither self-adjoint nor
unitary, consider
\begin{eqnarray}
{\rm diag}(2,i)
&=&
{\bf 1}+\sigma_3 +{i\over 2}({\bf 1} - \sigma_3) \nonumber \\
&=&
\left[
{1+i\over 2} {\bf 1}
+
{1-i\over 2} \sigma_3
\right]
\left[
{3\over 2} {\bf 1}
+
{1\over 2} \sigma_3
\right],
\end{eqnarray}
where $\sigma_3={\rm diag}(1,-1)$ and
both summands and factors commute and thus are co-measurable.


Co-measurability
is an important issue in the theory of partial algebras \cite{kochen3,kochen1},
where, in accordance with quantum mechanics,
operations are only allowed between mutually commuting operators corresponding to
co-measurable observables.
In particular, let us define the {\em context}
as the set of all co-measurable properties of a physical system.
By a well-known theorem, any context has associated with it a
single (though not unique) observable represented by a self-adjoint operator $C$
such that all other observables represented by self-adjoint $A_i$
within a given context
are merely functions (in finite dimensions polynomials) $A_i=f_i(C)$
thereof. We shall call $C$ the {\em context operator}.
Context operators are maximal in the sense that they exhaust their context
but they are not unique, since any one-to-one transformation of $C$
such as an isometry yields a context operator as well.

Different operators $A_i$ may belong to different contexts.
Actually, the proof of Kochen and Specker \cite{kochen1}
(of the nonexistence of consistent global truth values by associating such valuations
locally) is based on a finite chain of contexts linked together at one
operator per junction which belongs to the two contexts forming that junction.
This fact suggests that---rather than considering single operators
which may belong to different contexts---it is more appropriate
to  consider context operators instead. By definition, they carry the
entire context and thus cannot belong to different ones.
A graphical representation of context operators has been given
by Tkadlec \cite{tkadlec-96}, who suggested to consider dual Greechie diagrams
which represent context operators as vertices and links between different contexts
by edges.
A typical application would be the measurement of all the $N$ contexts
necessary for a Kochen-Specker contradiction
in an entangle $N$ particle singlet state.
In such a case, there should exist at least one observable
belonging to two different contexts whose outcomes are different
(cf also \cite{hey-red} for a similar reasoning).

\section{Pitowsky polytopes for automaton logics and
generalized urn models}

Let us assume that the chances of sunshine in Vienna $P_s(V)$ as well as in Budapest $P_s(B)$ are 50:50.
Would you believe a statement claiming that the joint
probability $P_s(A\wedge B)$ that the sun is shining in Budapest as well as in Vienna is 0.99?
No, I guess, you would not, since it appears unreasonable to claim that $P_s(A),P_s(B)\le P_s(A\wedge B)$.

In the middle of the 19th century the English mathematician George Boole
formulated a theory of "conditions of possible experience"
\cite{Boole,Boole-62,Hailperin}.
These conditions are related to relative frequencies of
logically connected
events and are expressed by certain equations or inequalities.
More recently, similar equations for a particular setup which are relevant in the
quantum mechanical context have been discussed by Bell, Clauser and  Horne and others
\cite{bell-87,cl-horne,chsh,clauser}.
Itamar Pitowsky has given a geometrical interpretation of classical
Boole-Bell  "conditions of possible experience" in terms of correlation polytopes
\cite{pitowsky-89a,pitowsky,Pit-91,Pit-94,2001-cddif}:
Take the probabilities $P_1, \ldots ,P_n$ of some events
$1,2,\ldots n$ and some (or all)
of the joint probabilities $P_1\wedge P_2,  \ldots ,P_{n-1}\wedge P_n,
P_1\wedge P_2\wedge P_3,  \ldots  $ and write them in vector form
${\bf x}=(P_1, \ldots ,P_n,P_1\wedge P_2,  \ldots ,P_{n-1}\wedge P_n,
P_1\wedge P_2\wedge P_3,  \ldots)$.
Every possible combination of all valuations\footnote{
In what follows, the terms ``two-valued (probability) measure'', ``two-valued state'',
``valuation'', and ``dispersion-free measure (state)'' will be used synonymously
for a lattice homomorphism $P:L\rightarrow {0,1}$ such that $P(L)=1$. }
of the $n$ Boolean algebras  $2^1$
formed by the atoms $\{i,i'\}$, $i=1,\ldots ,n$
($p'$ stands for the complement of $p$)
corresponds to one of the $2^n$ vertices (i.e., extreme points)
of a classical correlation polytope
\begin{equation}
\left\{\lambda_1 {\bf x}_1+\cdots +\lambda_l {\bf x}_l \;\Big|\;
l=2^n\ge 1, \lambda_j\ge 0, \sum_{j=1}^{2^n} \lambda_j=1\right\}
,
\label{e-cs}
\end{equation}
where ${\bf x}_i$ stands for the truth function corresponding to the $i$th valuation.
Thus, the vector components of ${\bf x}_j$ are either $0$ or $1$,
and the first $n$ components contain all $2^n$ possible distinct combinations thereof.

Every convex polytope in an Euclidean space has a dual description:
either as the convex hull of its vertices as in Eq. (\ref{e-cs}) (V-representation),
or as the intersection of a finite number of half-spaces,
each one given by a linear inequality (H-representation)
This equivalence is known as the \emph{Weyl-Minkowski} theorem (e.g., \cite[p. 29]{ziegler}).
The problem to obtain all inequalities from the vertices of a convex
polytope is known as the \emph{hull problem}. One solution strategy
is the Double Description Method \cite{MRTT53} which we shall use but not review here.

What Boole did not foresee, however, is that certain events
in one and the same inequality may be operationally incompatible,
and that the event structure may not be a Boolean algebra.
This applies to quantum mechanics as well as to Wright's generalized urn models \cite{wright:pent,wright},
as well as to automaton partition logics \cite{svozil-93,schaller-96,svozil-ql}.

The importance of correlation polytopes lies in the fact that they fully exploit
all consistently conceivable probabilities. Their border faces correspond to Boole-Bell type
inequalities. If dispersion-free states exist, every vertex corresponds to a dispersion-free state.
In this sense, correlation polytopes define the probabilities of a given formal structure
completely.

Classical correlation polytopes corresponding to important
quantum cases have already been studied
intensively  \cite{pitowsky,Pit-91} (for a recent investigation, see \cite{2000-poly,2001-cddif}).
Nonclassical correlation polytopes and thus the associated probabilities are less known
\cite{cirelson:80,cirelson:87,cirelson}.
In a different context, Ron Wright has investigated
states on nonclassical event structures in detail
\cite{wright:pent}.
In what follows, we shall use his analysis
to define Pitowsky correlation polytopes of automaton partition logics and
generalized urn models.

Whereas for Hilbert logics  of Hilbert spaces
with dimension higher than or equal to three,
no dispersion-free state exists \cite{kochen1},
for generalized urn logics and automaton partition logics,
two-valued states exist and can be used for an explicit construction of the respective models.
Just as for Boolean algebras  every probability can be composed by
the convex combination of two-valued states,
any probability on orthologics
(a bounded, orthocomplemented poset in which orthocomplemented joins exist)
(admitting a two-valued state)
is a convex combination of two-valued states
\cite[Theorems 1.6, 1.7]{wright:pent}.
We shall restrict our attention to orthologics $L$ with a separating set of states; i.e.,
for every $s,t\in L$ with $s\neq t$, there is a two-valued state $P$ such that $P(s)\neq P(t)$
(an even weaker criterion would be unitality).

One immediate question is the following one:
how do such nonclassical correlation polytopes
of orthologics admitting two-valued states relate to classical correlation polytopes?
An answer can be given
in analogy to the Boolean case:
The nonclassical correlation polytope ${\cal C}(L)$ corresponding to
some nonboolean lattice $L$ can be defined as the convex hull
of all two-valued states thereon.
That is, Eq. (\ref{e-cs}) also applies for the nonclassical case; with
the generalization that the set of vectors ${\bf x}_i$ corresponds to the set of all
two-valued states thereon.
(In the  quantum mechanical case, no valuations exist for Hilbert spaces of dimension bigger than two;
thus the definition cannot be applied to quantum correlation polytopes.)

Separability (unitality) implies embedability of a the orthologic $L$ into
a Boolean algebra $B=2^n$ with  $n$ atoms.
We may consider the corresponding correlation polytope  ${\cal C}_n$ generated by
the subset of its $2^n$ vertices, which are its extreme points if the truth
assignments are identified by vector components.
The dimension of the vector space depends on the number of propositions involved.
Since not all valuations of the Boolean algebra $2^n$
need to be valuations of $L$, ${\cal C}(L)$ is  a subset of  ${\cal C}_n$.

It should also be noted that the requirement that the automaton system can only be
in a single one of the states 1,2 and 3 imposes additional restrictions
and effectively reduces the number of vertices.
Let us consider a very simple explicit example.
Consider the automaton partition logic
$$L=\{
\{
\{
1
\}
\{
2,3
\}
\}
.
\{
\{
2
\}
\{
1,3
\}
\}
.
\{
\{
3
\}
\{
1,2
\}
\}
\}.
$$

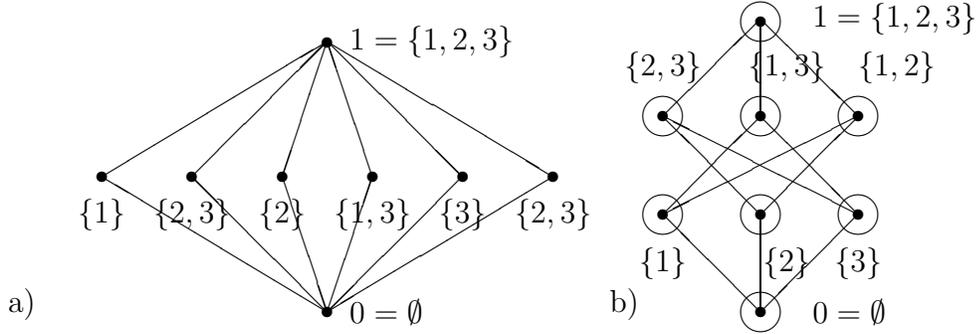
\begin{figure}
\begin{center}
a)
\unitlength 0.60mm
\linethickness{0.4pt}
\begin{picture}(111.06,60.73)
\put(60.00,0.00){\circle*{2.11}}
\put(30.00,30.00){\circle*{2.11}}
\put(60.00,59.67){\circle*{2.11}}
\put(90.00,30.00){\circle*{2.11}}
\put(60.00,0.00){\line(-1,1){30.00}}
\put(30.00,30.00){\line(1,1){30.00}}
\put(60.00,60.00){\line(1,-1){30.00}}
\put(90.00,30.00){\line(-1,-1){30.00}}
\put(65.00,0.00){\makebox(0,0)[lc]{$0=\emptyset$}}
\put(65.00,60.00){\makebox(0,0)[lc]{$1=\{1,2,3\}$}}
\put(30.00,25.00){\makebox(0,0)[ct]{$\{2,3\}$}}
\put(90.00,25.00){\makebox(0,0)[ct]{$\{3\}$}}
\put(60.00,0.00){\line(-5,3){50.00}}
\put(10.00,30.00){\line(5,3){50.00}}
\put(60.00,60.00){\line(5,-3){50.00}}
\put(110.00,30.00){\line(-5,-3){50.00}}
\put(10.00,30.00){\circle*{2.11}}
\put(10.00,25.00){\makebox(0,0)[ct]{$\{1\}$}}
\put(110.00,30.00){\circle*{2.11}}
\put(110.00,25.00){\makebox(0,0)[ct]{$\{2,3\}$}}
\put(70.00,30.00){\circle*{2.11}}
\put(70.00,25.00){\makebox(0,0)[ct]{$\{1,3\}$}}
\put(50.00,30.00){\circle*{2.11}}
\put(50.00,25.00){\makebox(0,0)[ct]{$\{2\}$}}
\put(60.00,60.00){\line(-1,-3){10.00}}
\put(50.00,30.00){\line(1,-3){10.00}}
\put(60.00,0.00){\line(1,3){10.00}}
\put(70.00,30.00){\line(-1,3){10.00}}
\end{picture}
$\quad$
b)
\unitlength 1.30mm
\linethickness{0.4pt}
\begin{picture}(22.33,31.67)
\put(0.33,10.00){\circle*{1.00}}
\put(10.33,10.00){\circle*{1.00}}
\put(20.33,10.00){\circle*{1.00}}
\put(0.33,20.00){\circle*{1.00}}
\put(10.33,20.00){\circle*{1.00}}
\put(20.33,20.00){\circle*{1.00}}
\put(10.33,29.67){\circle*{1.00}}
\put(10.33,0.00){\circle*{1.00}}
\put(10.33,0.00){\line(-1,1){10.00}}
\put(0.33,10.00){\line(1,1){10.00}}
\put(10.33,20.00){\line(0,1){10.00}}
\put(10.33,30.00){\line(1,-1){10.00}}
\put(20.33,20.00){\line(-2,-1){20.00}}
\put(10.33,0.00){\line(0,1){10.00}}
\put(10.33,10.00){\line(-1,1){10.00}}
\put(0.33,20.00){\line(1,1){10.00}}
\put(10.33,10.00){\line(1,1){10.33}}
\put(10.33,20.00){\line(1,-1){10.00}}
\put(20.33,10.00){\line(-2,1){20.00}}
\put(20.33,10.00){\line(-1,-1){10.00}}
\put(0.33,5.00){\makebox(0,0)[cc]{$\{1\}$}}
\put(20.33,5.00){\makebox(0,0)[cc]{$\{3\}$}}
\put(15.66,0.00){\makebox(0,0)[lc]{$0=\emptyset$}}
\put(0.33,25.00){\makebox(0,0)[cc]{$\{2,3\}$}}
\put(20.33,25.00){\makebox(0,0)[lc]{$\{1,2\}$}}
\put(15.66,30.00){\makebox(0,0)[lc]{$1=\{1,2,3\}$}}
\put(0.33,10.00){\circle{4.00}}
\put(20.33,10.00){\circle{4.00}}
\put(0.33,20.00){\circle{4.00}}
\put(20.33,20.00){\circle{4.00}}
\put(10.33,0.00){\circle{4.00}}
\put(10.33,29.67){\circle{4.00}}
\put(10.33,10.00){\circle{4.00}}
\put(10.33,20.00){\circle{4.00}}
\put(13.00,5.00){\makebox(0,0)[cc]{$\{2\}$}}
\put(13.00,25.00){\makebox(0,0)[cc]{$\{1,3\}$}}
\end{picture}
\end{center}
\caption{\label{2001-cesena-f1} Hasse diagram of a (surjective) embedding of
$MO_3$ drawn in a)
into $2^3$ drawn in b).
Concentric circles indicate points of $2^3$ included in
$MO_3$.}
\end{figure}

$L=MO_3$ is embedable into $B=2^3$ with the set of atoms $\{1,2,3\}$
in a straightforward manner by identifying the automaton states $1, 2, 3$
with these atoms; cf. Fig. \ref{2001-cesena-f1}.
The  correlation polytope ${\cal C}(L)$ of $L=MO_3$ consists of 3 vertices
$(1,0,0)$, $(0,1,0)$ and $(0,0,1)$.
The corresponding (in)equalities can be easily obtained by solving the hull
problem:
$P_1,P_2,P_3\ge 0$,
$P_1+P_2+P_3=1$.
Since all automaton logics have a set theoretic embedding of the above kind,
their corresponding correlation polytopes are subsets of the correlation polytopes of the
classical algebras in which they can be embedded.

\begin{figure}
\begin{center}
\unitlength 0.85mm
\linethickness{0.4pt}
\begin{picture}(108.00,55.00)
\put(15.00,17.09){\circle{2.00}}
\put(25.00,7.33){\circle{2.00}}
\put(55.00,27.33){\circle{2.00}}
\put(85.00,7.33){\circle{2.00}}
\put(95.00,17.33){\circle{2.00}}
\put(25.00,7.33){\line(1,0){60.00}}
\put(25.00,47.33){\line(1,0){60.00}}
\put(55.00,7.33){\line(0,1){40.00}}
\put(25.00,7.33){\line(-1,1){20.00}}
\put(5.00,27.33){\line(1,1){20.00}}
\put(85.00,7.33){\line(1,1){20.00}}
\put(105.00,27.33){\line(-1,1){20.00}}
\put(24.67,55.00){\makebox(0,0)[rc]{$a_3=\{10,11,12,13,14\}$}}
\put(55.33,55.00){\makebox(0,0)[cc]{$a_4=\{2,6,7,8\}$}}
\put(85.33,55.00){\makebox(0,0)[lc]{$a_5=\{1,3,4,5,9\}$}}
\put(9.00,40.00){\makebox(0,0)[rc]{$a_2=\{4,5,6,7,8,9\}$}}
\put(99.33,40.00){\makebox(0,0)[lc]{$a_6=\{2,6,8,11,12,14\}$}}
\put(0.00,26.33){\makebox(0,0)[rc]{$a_1=\{1,2,3\}$}}
\put(108.00,26.33){\makebox(0,0)[lc]{$a_7=\{7,10,13\}$}}
\put(60.33,31.33){\makebox(0,0)[lc]{$a_{13}=$}}
\put(60.33,26.33){\makebox(0,0)[lc]{$\{1,4,5,10,11,12\}$}}
\put(9.00,13.33){\makebox(0,0)[rc]{$a_{12}=\{4,6,9,12,13,14\}$}}
\put(99.67,13.33){\makebox(0,0)[lc]{$a_8=\{3,5,8,9,11,14\}$}}
\put(24.67,-0.05){\makebox(0,0)[rc]{$a_{11}=\{5,7,8,10,11\}$}}
\put(55.33,-0.05){\makebox(0,0)[cc]{$a_{10}=\{3,9,13,14\}$}}
\put(85.33,-0.05){\makebox(0,0)[lc]{$a_9=\{1,2,4,6,12\}$}}
\put(5.00,27.33){\circle{2.00}}
\put(15.00,37.33){\circle{2.00}}
\put(25.00,47.33){\circle{2.00}}
\put(55.00,47.33){\circle{2.00}}
\put(85.00,47.33){\circle{2.00}}
\put(55.00,7.33){\circle{2.00}}
\put(104.76,27.33){\circle{2.00}}
\put(95.00,37.33){\circle{2.00}}
\end{picture}
\end{center}
\caption{\label{2001-cesena-f2} Greechie diagram of automaton partition logic
with a nonfull set of dispersion-free measures.}
\end{figure}
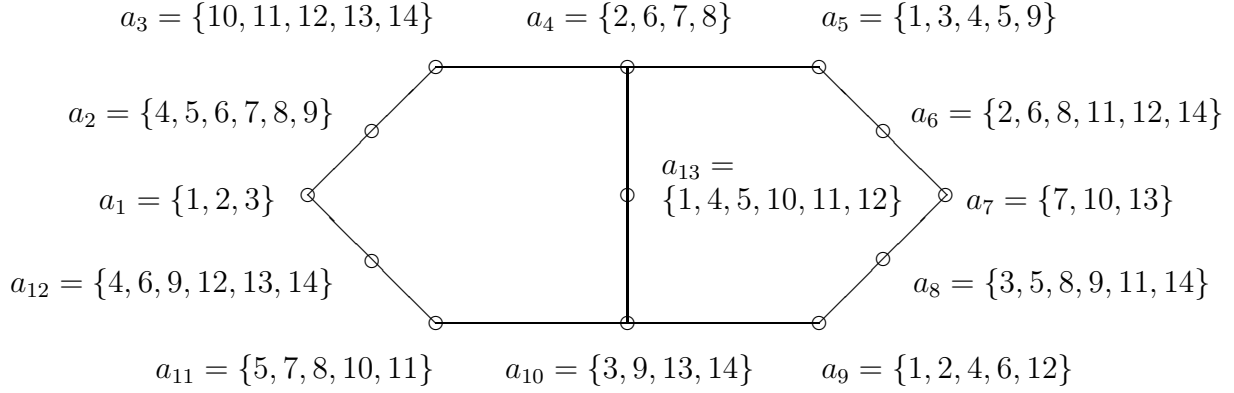

Another, less trivial example, is a logic which is already mentioned by Kochen
and Specker \cite{kochen1} (this is a subgraph of their $\Gamma_1$)
whose automaton partition logic is depicted in Fig. \ref{2001-cesena-f2}.
The  correlation polytope of this lattice consists of 14 vertices listed
in Table \ref{2001-cesena-t2}, where  the 14 rows indicate the vertices
corresponding to the 14 dispersion-free states. The columns
represent the partitioning of the automaton states.
\begin{table}
\caption{
Truth table of a logic with 14 dispersion-free states.
The rows, interpreted as vectors, are just the vertices of the corresponding correlation polytope in 13 dimensions.
\label{2001-cesena-t2}}
\begin{center}
\begin{tabular}{c|cccccccccccccccccccccccccc}
\hline\hline
\#&$a_1$&$a_2$&$a_3$&$a_4$&$a_5$&$a_6$&$a_7$&$a_8$&$a_9$&$a_{10}$&$a_{11}$&$a_{12}$&$a_{13}$&$a_{1}\wedge a_{2}$ &$\cdots$\\
\hline
1  &1&0&0&0&1&0&0&0&1&0&0&0&1&0     \\
2  &1&0&0&1&0&1&0&0&1&0&0&0&0&0     \\
3  &1&0&0&0&1&0&0&1&0&1&0&0&0&0     \\
4  &0&1&0&0&1&0&0&0&1&0&0&1&1&0     \\
5  &0&1&0&0&1&0&0&1&0&0&1&0&1&0     \\
6  &0&1&0&1&0&1&0&0&1&0&0&1&0&0     \\
7  &0&1&0&1&0&0&1&0&0&0&1&0&0&0     \\
8  &0&1&0&1&0&1&0&1&0&0&1&0&0&0     \\
9  &0&1&0&0&1&0&0&1&0&1&0&1&0&0     \\
10 &0&0&1&0&0&0&1&0&0&0&1&0&1&0     \\
11 &0&0&1&0&0&1&0&1&0&0&1&0&1&0     \\
12 &0&0&1&0&0&1&0&0&1&0&0&1&1&0     \\
13 &0&0&1&0&0&0&1&0&0&1&0&1&0&0     \\
14 &0&0&1&0&0&1&0&1&0&1&0&1&0&0     \\
\hline\hline
\end{tabular}
\end{center}
\end{table}
The solution of the hull problem by
the {\tt LPoly} package due to Maximian Kreuzer
and Harald Skarke \cite{kreuzer-skarke} yields the equalities
\cite{kreuzer-priv}
\begin{equation}
\label{2001-cesena-eq18}
\begin{array}{l}
1 =   P_1  +P_2  +P_3 =   P_4 +P_{10} +P_{13}                          ,\\
1 = P_1+P_2 -P_4  +P_6+P_7 =  -P_2+P_4 -P_6+P_8 -P_{10}+P_{12}                  ,\\
1 =   P_1  +P_2 -P_4+P_6-P_8+P_{10}  +P_{11} ,\\
0 = P_1  +P_2-P_4 -P_5  = -P_1-P_2  +P_4 -P_6  +P_8  +P_9                    ,\
.
\end{array}
\end{equation}
The operational meaning of $P_i=P_{a_i}$  is
``the probability to find the automaton in state $a_i$.''
Eqs. (\ref{2001-cesena-eq18}) are equivalent to all probabilistic conditions on the contexts
(subalgebras)
$1
=P_1+P_2+P_3
=P_3+P_4+P_5
=P_5+P_6+P_7
=P_7+P_8+P_9
=P_9+P_{10}+P_{11}
=P_4+P_{10}+P_{13}
$.

Let us now turn to the joint probability case.
Notice that formally it is possible to form a statement such as $a_1\wedge a_{13}$
(which would be true for measure number $1$ and false otherwise),
but this is not operational on a single automaton subject to the
Moore measurement conditions  \cite{e-f-moore},
since no experiment can decide such a proposition on a single automaton.
Nevertheless, if one considers a a ``singlet state'' of two automata which are in an unknown
yet identical initial state, then an expression such as $a_1\wedge a_{13}$ makes operational sense
if property $a_1$ is measured on the first automaton
and property $a_{13}$ on the second automaton. Indeed, all joint probabilities
$a_i\wedge a_j\wedge \ldots a_n$
make sense  in an $n$-automaton singlet context.

\section{Parameter cheats}

In this section, certain bijective (one-to-one) parameter  transformations will be performed which
artificially give classical systems a quantum flavor; and conversely,
seemingly make quantum systems behave classically,
at least with respect to joint probabilities.
Since such transformations have other, undesirable features, we shall call them
``parameter cheats.''

\begin{figure}
\begin{center}
 \includegraphics[width=10cm]{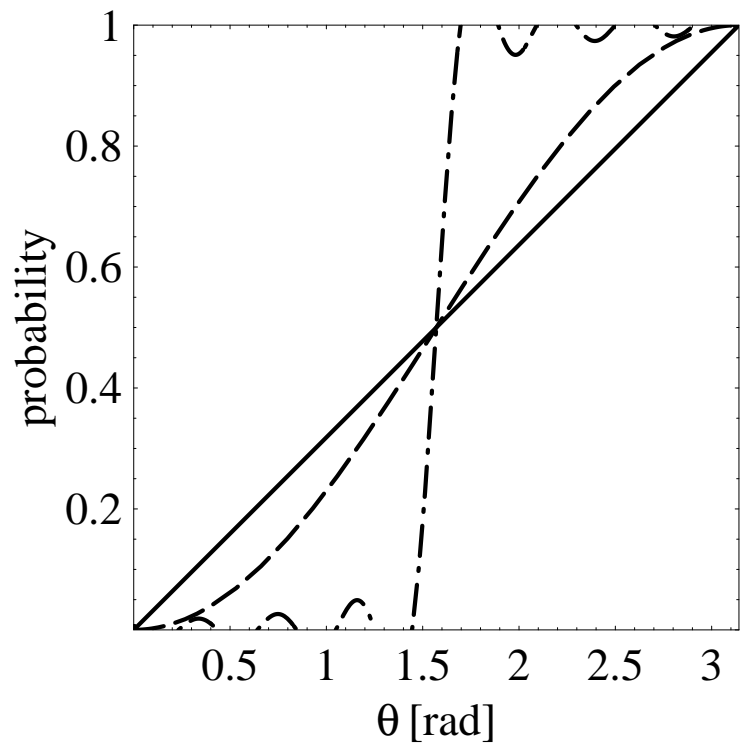}
\end{center}
 \caption{
Different joint probability functions of the  parameter $\theta$.
The solid, dashed and dot-dashed lines indicate classical, quantum and
stronger-than-quantum behavior ($n=11$), respectively.
}
\label{2001-cheat-fprob}
\end{figure}

Consider a singlet state
for which the sum of all angular momenta and spins is zero.
In the quantum mechanical case, let us assume two particles of spin 1/2
in an EPR-Bohm configuration.
Then the probability
$P^{=}(\theta )$
to find the angular momentum or spin of
both particles
measured along two axis which are an angle $\theta $ apart
in the same direction is given by \cite{svozil-krenn}

\begin{eqnarray}
P^{=}_{qm} (\theta )&=&\sin^2 (\theta /2)\\
P^{=}_{cl} (\theta ) &=& \theta /\pi       \\
P^{=}_{s} (\theta ) &=&
{1\over 2}
+
{2\over \pi}
\sum_{k=0}^{n>1}
{
\sin \left[(2k+1)\left({2\Delta/ \pi}-1\right) \right]
\over 2k+1}
\nonumber \\
&\stackrel{n\rightarrow \infty}{\longrightarrow}& H(2\theta /\pi -1)=(1/2)(1+{\rm sgn} (2\theta /\pi
-1))
\end{eqnarray}
for $0\le \theta \le \pi$.
$P^{=}_{qm} (\theta )$,
$P^{=}_{cl} (\theta ) $,
$P^{=}_{s} (\theta ) $ stand for the joint classical, quantum and
stronger-than-quantum probabilities, respectively.
Figure   \ref{2001-cheat-fprob} represents different joint probability functions of the  parameter $\theta$.

\subsection{Quantum cheat for classical system}

\begin{figure}
\begin{center}
 \includegraphics[width=10cm]{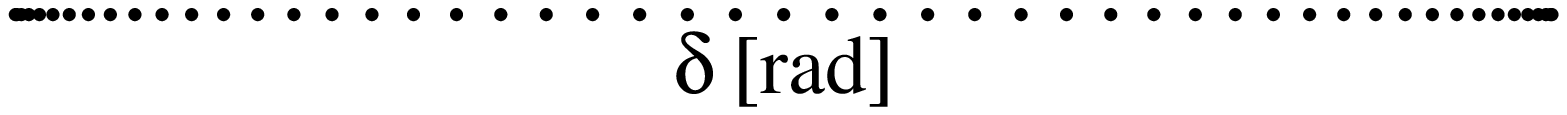}
\end{center}
\begin{center}a)\end{center}
\begin{center}
 \includegraphics[width=10cm]{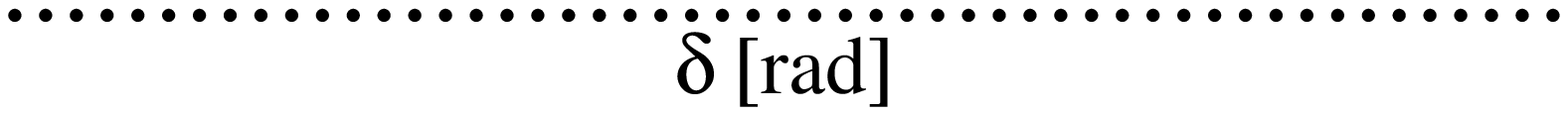}
\end{center}
\begin{center}b)\end{center}
 \caption{
a) Evaluation of the deformed parameter scale $\delta$ versus $\theta$.
b) Evaluation of the linear reference parameter $\delta$.
}
\label{2001-cheat-f1}
\end{figure}

Then, in order to be able to fake
a quantum form of the classical expression,
we introduce a ``cheat parameter'' $\delta $,
which is obtained from the angular parameter $\theta $
by a nonlinear transformation
$T:\theta \mapsto \delta$
from the Ansatz
\begin{equation}
P^{=}_{cl}(\theta (\delta ))
  =
P^{=}_{cl}(\delta )
=
{\theta (\delta )\over \pi}=
\sin^2 \left({\delta \over 2}\right)
.
\label{2001-e1}
\end{equation}
The right hand side of Eq. (\ref{2001-e1})
yields
\begin{eqnarray}
\theta &=& \pi \sin^2 \left({\delta \over 2} \right)\\
\delta &=& 2\arcsin \sqrt{{\theta \over \pi}}
\end{eqnarray}
where $0\le \delta  \le \pi$.
Figure   \ref{2001-cheat-f1} represents a numerical evaluation
of the deformed parameter scale $\delta$ in terms $\theta$.

\subsection{Classical cheat for quantum system}

\begin{figure}
\begin{center}
 \includegraphics[width=10cm]{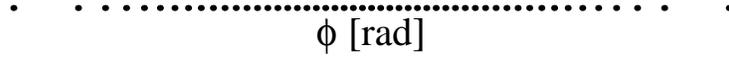}
\end{center}
 \caption{Evaluation of the deformed parameter scale $\phi$ versus $\theta$.}
\label{2001-cheat-f2}
\end{figure}

In order to be able to fake
a classical form of the quantum expression,
we introduce a ``cheat parameter'' $\phi $,
which is obtained from the angular parameter $\theta $
by a nonlinear transformation
$T:\theta \mapsto \phi$
from the Ansatz
\begin{equation}
P^{=}_{qm}(\theta (\phi ))
=
P^{=}_{qm}(\phi )
=
{\phi \over \pi}=
\sin^2 \left({\theta (\phi ) \over 2}\right).
\label{2001-e2}
\end{equation}
The right hand side of Eq. (\ref{2001-e2})
yields
\begin{eqnarray}
\theta &=& 2\arcsin \sqrt{{\phi \over \pi}}\\
\phi &=& \pi \sin^2 \left({\theta \over 2} \right)
\end{eqnarray}
where $0\le \phi  \le \pi$.
Figure   \ref{2001-cheat-f2} represents a numerical evaluation
of the deformed parameter scale $\phi$ in terms $\theta$.

\subsection{Stronger-than-quantum (STQ) cheat for classical system}

In order to be able to fake
a STR form of the classical expression,
we introduce a ``cheat parameter'' $\Delta $,
which is obtained from the angular parameter $\theta $
by a nonlinear transformation
$T:\theta \mapsto \Delta$
from the Ansatz
\begin{eqnarray}
P^{=}_{cl}(\theta (\Delta ))
&=&
\nonumber \\
P^{=}_{cl}(\Delta )
&=&
{1\over 2}
+
{2\over \pi}
\sum_{k=0}^n
{
\sin \left[(2k+1)\left({2\Delta/ \pi}-1\right) \right]
\over 2k+1}
=
{\delta (\Delta )\over \pi }
,
\label{2001-e3}
\end{eqnarray}
where $n\ge 1$.
In the limit,
$$\lim_{n\rightarrow \infty}
{4\over \pi}
\sum_{k=0}^n
{
\sin \left[(2k+1)\left({2\Delta \over \pi}-1\right) \right]
\over 2k+1}
=
{\rm sgn}\left({2\Delta \over \pi}-1\right).
$$

The right hand side of Eq. (\ref{2001-e3})
yields
\begin{eqnarray}
\theta &=&
{\pi \over 2}
+
{2}
\sum_{k=0}^n
{
\sin \left[(2k+1)\left({2\Delta/ \pi}-1\right) \right]
\over 2k+1}
\end{eqnarray}

\subsection{How do the cheats perform?}

Cheats perform in a very simple way, which can be best understood if one considers
the ``proper'' physical parameter  and compares it to the ``cheat'' parameter.
The cheat parameter effectively deforms the proper parameter range in that
measures therein pretend to be in a different parameter range than the one in which
the proper parameter is. It is quite clear then, that cheats can mimic almost any
behavior as long as the parameter transformation remains  one-to-one.

Let us consider the Clauser-Horne (CH) inequality
\begin{equation}
-1\leq P(A_{1}B_{1})+P(A_{1}B_{2})+P(A_{2}B_{2})-P(A_{2}B_{1})-P(A_{1})-P(B_{2}) \leq 0
\label{2001-cheat-ech}
\end{equation}
and a classical system on which a quantum cheat has been applied.
Let the angles be
\begin{eqnarray}
A_{1}&:&\delta_1 = 0,
\nonumber \\
B_{1}&:&\delta_2 = \pi /4,
\nonumber \\
A_{2}&:&\delta_3 = \pi /2,
\nonumber \\
B_{2}&:&\delta_4 = 3\pi /4.
\nonumber
\end{eqnarray}

Identify
$P(A_{i})=P(B_{i})= 1/2$
and
\begin{eqnarray}
P(A_1B_1)&=& P^{=}_{cl}((\delta_2 -\delta_1)/2 = \pi /8),
\nonumber \\
P(A_2B_2)&=& P^{=}_{cl}((\delta_4 -\delta_3)/2 = \pi /8),
\nonumber \\
P(A_1B_2)&=& P^{=}_{cl}((\delta_4 -\delta_1)/2 =3\pi /8),
\nonumber \\
P(A_2B_1)&=& P^{=}_{cl}((\delta_3 -\delta_2)/2 = \pi /8).
\nonumber
\end{eqnarray}
With a choice of these angles, the right hand side of Eq. (\ref{2001-cheat-ech}) is violated.

Of course, we cannot expect from the cheat parameter to inherit the linear behavior of the
old parameter; in particular
$
\delta_3(\theta_3)
=
\delta_1(\theta_1)+
\delta_2(\theta_2)$ does not imply $\theta_3=\theta_1+\theta_2$,
and
$\delta(\theta_1)+\delta(\theta_2)=\delta(\theta_3)$; i.e.,
\begin{equation}
\delta_3(\theta_1+\theta_2)
\neq
\delta_1(\theta_1)+
\delta_2(\theta_2)
\end{equation}
Therefore, one might call a parameter description to be ``proper''
if it is isotropic and linear with respect to a reference scheme.
Of course, this leaves open the question whether or not it makes sense to
refer to particular parameter descriptions as absolute ones;
yet at least in the typical experimental physical context this seems evident
and appropriate for most physical purposes.
In such a scheme, the above mentioned cheat parameters are improper.

\section{Summary}

There exist extensions of quantum mechanics
guided by Hilbert space theory
which may be considered
as generalizations of the standard formalism.
All these extensions are operationalizable and may thus contribute
to a better understanding of the quantum phenomena.

In the second section we discussed Boole-Bell type inequalities
and Pitowsky correlation polytopes as a criterion for probabilities
of automaton partition logics and generalized urn models.
We find that, although the event structure is nonboolean,
the corresponding probabilities can be represented as linear combinations of
dispersion-free states and thus by the hull of the vertices defined by them.
The corresponding correlation polytope is a subset of the classical correlation
polytope of the Boolean algebra in which these logics can be embedded.
The issue of Pitowsky correlation polytopes which exceed their classical
counterpart remains open.

Finally, parameter transformation have been discussed which translate classical
correlations into nonclassical ones and {\it vice versa}.
While at the first glance the possibility for such a representation seems counterintuitive,
a more detailed analysis reveals that the corresponding parameters
can be used consistently but have undesirable
features such as nonuniformity.

\section*{Acknowledgments}
The author gratefully acknowledges many stimulating Friday discussion with
Johann Summhammer and G{\"{u}}nther Krenn in the
Viennese coffee house ``Prueckl.''

\end{document}